\documentclass[runningheads]{llncs}

\usepackage{graphicx}
\usepackage{ulem}
\usepackage{amsmath}
\usepackage{amssymb}
\usepackage{xcolor}
\usepackage{float}
\usepackage{booktabs}

\begin{document}

\title{Approximate Knowledge Graph Query Answering:\\  From Ranking to Binary Classification
}

\titlerunning{KG Query Answering with Binary Classification}

\author{Ruud van Bakel\inst{2,1}\orcidID{0000-0002-6891-5237} \and
Teodor Aleksiev\inst{3,1}\orcidID{0000-0002-6567-5960} \and
Daniel Daza\inst{1,2,4}\orcidID{0000-0002-5357-3705} \and
Dimitrios Alivanistos\inst{1,4}\orcidID{0000-0002-0090-2069} \and
Michael Cochez\inst{1,4}\orcidID{0000-0001-5726-4638}
}

\authorrunning{van Bakel, Aleksiev, et al.}

\institute{
Computer Science, Vrije Universiteit Amsterdam, The Netherlands \\
\email{\{d.dazacruz, d.alivanistos, m.cochez\}@vu.nl}\\
\and
University of Amsterdam, The Netherlands
\email{ruudvanbakel@yahoo.co.uk} 
\and
Leiden University, The Netherlands
\email{aleksiev.teodord@gmail.com}
\and
Discovery Lab, Elsevier, The Netherlands
\email{https://discoverylab.ai}
}

\maketitle              

\begin{abstract}

Large, heterogeneous datasets are characterized by missing or even erroneous information. This is more evident when they are the product of community effort or automatic fact extraction methods from external sources, such as text.
A special case of the aforementioned phenomenon can be seen in knowledge graphs, where this mostly appears in the form of missing or incorrect edges and nodes.

Structured querying on such incomplete graphs will result in incomplete sets of answers, even if the correct entities exist in the graph, since one or more edges needed to match the pattern are missing.
To overcome this problem, several algorithms for approximate structured query answering have been proposed.
Inspired by modern Information Retrieval metrics, these algorithms produce a ranking of all entities in the graph, and their performance is further evaluated  based on how high in this ranking the correct answers appear.

In this work we take a critical look at this way of evaluation. 
We argue that performing a ranking-based evaluation is not sufficient to assess methods for complex query answering.
To solve this, we introduce Message Passing Query Boxes (MPQB), which takes binary classification metrics back into use and shows the effect this has on the recently proposed query embedding method MPQE.

\keywords{Query answering  \and geometric representation \and Box embeddings \and approximation.}
\end{abstract}

\section{Introduction}

In many organizations, a vast amount of complex information is used in operations daily.
This data is often stored in various databases or file systems while information can be retrieved using query languages and information retrieval techniques. During the past decade, several companies have started taking up knowledge graphs (KG) \cite{hogan2020knowledge}, as a way to represent heterogeneous data and make it useful for a large variety of applications \cite{noy2019industry}.
To make said data accessible, various querying languages like SPARQL and Cypher have been developed. Such querying languages allow for accessing nodes in the graph, traversing them via specific relations, or retrieve nodes that match a specific pattern. At the core of these languages lie graph patterns. 
These patterns can be thought of as graph shaped structures where some nodes and edges can correspond to nodes existing in the graph, while others correspond to variables (with specific variable names).
When a match for this pattern is found in the graph, the variables are bound and the appropriate values are returned as the result.

However, the performance of the previously described process is heavily dependent on the level of completeness in the graph. 

To go in detail, completeness refers to whether it contains all the nodes and edges in the graph pattern, and has a binding for all variables. 
Having a single node or edge missing from the graph, which represents a comparatively small bit of information, results in missing answers.
This phenomenon could be good, in case of an erroneous piece of information, or bad, in case of information missing from the graph.

In this paper, we focus on this issue, specifically the case of missing edges in the graph.
Ideally, we would like a query system that can still give answers when the phenomenon described before applies. We would like to have \textit{approximate query answering}. 

One way to approach this, is by performing link prediction. 
In link prediction, one would try to predict missing links in the graph, by training a machine learning model on the known parts of it.
While not trivial, it is possible to use the single link prediction mechanism to answer queries with missing links.
Another way to approach this problem is by using the so-called query encoders. 
These encoders take a query as input and produce an embedding (a high dimensional vector representation) for it. 
This query embedding is later compared to learned embeddings for the entities in the graph.
This machine learning system is optimised in such a way that entities close to the query embedding in vector space, are also its probable answers.

In this paper we focus on the analysis and evaluation of these systems. 
Typically, such systems return a series of candidate answers to the query, accompanied by a likelihood or distance from the query embedding in vector space.
In the evaluation phase, this ranking is compared to, not a ground truth ranking, but rather the set of correct answers to the query.
To do this, typical measures like hits@n (how many correct answers out of n) and mean reciprocal rank (MRR -- what is the average reciprocal of the rank of correct answers) are used.
While these measures are appropriate for information retrieval systems, they fall short when it comes to query systems. In the latter, the results are not ranked, but are rather the correct answer or not.

This is also reflected in how these measures are usually adapted by modifying them to filtered versions.
In this case, measures like hits@n and MRR are computed such that true answers higher in the returned ranking are ignored when computing for example the rank for lower ranked entities.

We argue that we need to look into metrics that are not based on specific ranking of the results, but rather on a crisp set of results retrieved from these systems. 
A main argument for why this is necessary is that many downstream tasks using the aforementioned results need to get a finite set of answers from the knowledge graph, not just a ranked list of all possible entities.
That is, we need a query engine that does not just act as a ranking system, but as a binary classifier: it must provide a set of entities that are answers to the query while all other entities are not.
In this scenario, the evaluation would be the same as what has traditionally been used for classification problems, with measures such as precision and recall.

This paper is structured as follows:
in section 2, we provide an example for several algorithms used for approximate query answering.
Then, in section 3 we discuss how metrics for binary classification can provide additional insight on top of the metrics used for ranking. We end that section with a general direction on how this could be achieved in the existing systems using volumetric query embeddings.
Section 4 details a first approach for solving this problem using axis-aligned hyper-rectangles for these queries. We describe the MPQB model, a proof-of-concept, in the section after that.
Finally, we provide a conclusion and future outlook.

This work is largely based on the Bachelor thesis works of Ruud van Bakel~\cite{thesisRuud} and Teodor Aleksiev~\cite{thesisTeodor}, who both worked under the supervision of Michael Cochez at the Vrije Universiteit Amsterdam.

\section{Approximate Query Answering on Knowledge Graphs}

\label{sec:methods}

We define a knowledge graph as a tuple $\mathcal{G} = (\mathcal{V}, \mathcal{R}, \mathcal{E})$, where $\mathcal{V}$ is a set of entities, $\mathcal{R}$ a set of relation types, and $\mathcal{E}$ a set of binary predicates of the form $r(h, t)$ where $r\in\mathcal{R}$ and $h,t\in\mathcal{V}$. Each binary predicate represents an edge of type $r$ between the entities $h$ and $t$, and thus we call $\mathcal{E}$ the set of edges in the knowledge graph.

A query on a KG looks for the set of entities that meet a particular condition, specified in terms of binary predicates whose arguments can be constants (i.e. entities in $\mathcal{V}$), or variables.
As an example, consider the following query (adapted from \cite{daza2020message}): ``Select all projects $P$, such that topic $T$ is related to $P$, and both \textit{Alice} and \textit{Bob} work on $T$''. In this query, the constants entities are \textit{Alice} and \textit{Bob}, and the variables are denoted as $P$ and $T$. We can define such a query formally in terms of a conjunction of binary predicates, as follows:
\begin{equation}
    q = P.\exists T, P:~ \mathrm{related}(T, P) \wedge \mathrm{works\_on}(\text{Alice}, T) \wedge \mathrm{works\_on}(\text{Bob}, T).
\end{equation}

More formally, we are interested in answering \textit{conjunctive queries}, that have the following general form:
\begin{equation}
q = V_t.\exists V_1, \dots, V_m: r_1(a_1, b_1) \wedge \ldots \wedge r_m(a_m, b_m),
\label{eq:query_def}
\end{equation}
In this notation, $r_i\in\mathcal{R}$, and $a_i$ and $b_i$ are constant entities in the KG, or variables from the set $\lbrace V_t, V_1, \dots, V_m \rbrace$.

Recent works have proposed to use machine learning methods to answer such queries. These methods operate by learning a vector representation in a space $\mathbb{R}^d$ for each entity and relation type. These representations are also known as \textit{embeddings}, and we denote them as $\mathbf{e}_v$ for $v\in\mathcal{V}$ and $\mathbf{e}_r$ for $r\in\mathcal{R}$. Similarly, these methods define a \textit{query embedding function} $\phi$ (usually defined with some free parameters), that maps a query $q$ to an embedding $\phi(q) = \mathbf{q}\in\mathbb{R}^d$. 

Given a query embedding $\mathbf{q}$, a score for every entity in the graph can be obtained via cosine similarity:
$$
\mathrm{score}(\mathbf{q}, \mathbf{e}_v) = \frac{\mathbf{q}^\top\mathbf{e}_v}{\Vert \mathbf{q} \Vert \Vert \mathbf{e}_v \Vert}.
$$
The entity and relation type embeddings, as well as any free parameters in the embedding function $\phi$, are optimized via stochastic gradient descent on a specific loss function. Usually the loss is defined so that for a given embedding of a query, the cosine similarity is maximized with embeddings of entities that answer the query, and minimized for embeddings of entities sampled at random. 

The dataset used for training consists of query-answer pairs mined from the graph. Once the procedure terminates, the function $\phi$ can be used to embed a query. The entities in the graph can then be ranked as potential answers, by computing the cosine similarity of all the entity embeddings and the embedding of the query.

Note that in contrast with classical approaches to query answering, such as the use of SPARQL in a graph database, this approach can return answers even if no entity in the graph matches exactly every condition in the query.

In the next sections we review the specifics of recently proposed methods, which consider particular geometries for embedding entities, relation types, and queries; as well as scoring functions.

\subsection{GQE}

\begin{figure}[t]
    \centering
    \includegraphics[scale=0.4]{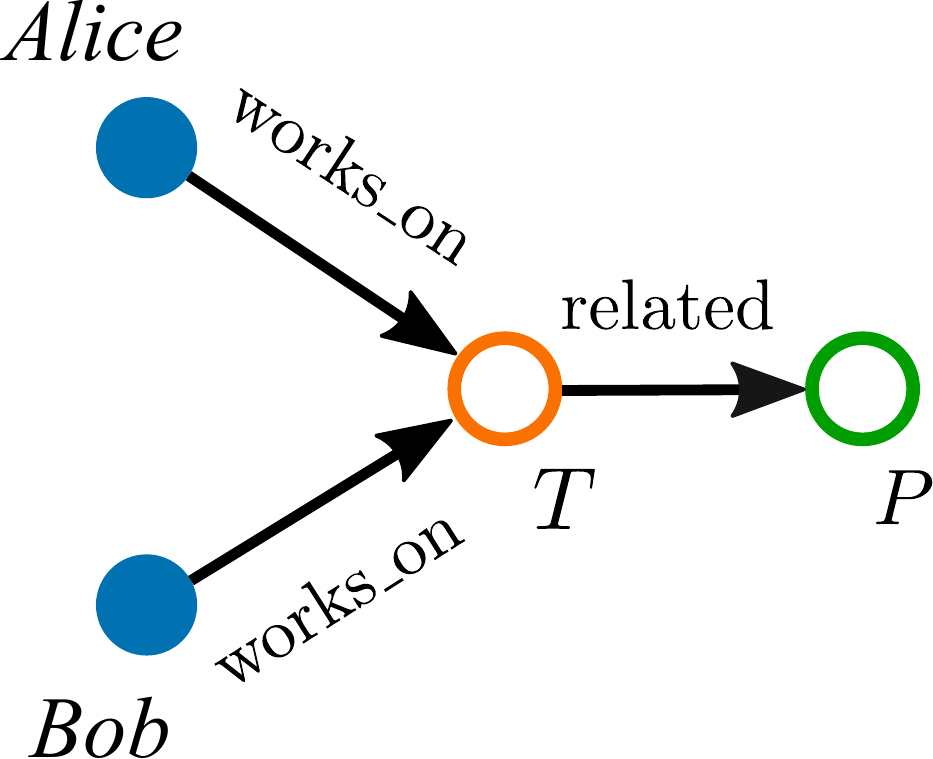}
    \caption{The query $q = P.\exists T, P:~ \mathrm{related}(T, P) \wedge \mathrm{works\_on}(\text{Alice}, T) \wedge \mathrm{works\_on}(\text{Bob}, T)$ can be represented as a directed acyclic graph, where the leaves are constant entities, the intermediate node $T$ is a variable, and $P$ is the target entity.  (adapted from a figure in \cite{daza2020message})}
    \label{fig:query-graph}
\end{figure}

Conjunctive queries can be represented as a directed acyclic graph, where the leaf nodes are constant entities, any intermediate nodes are variables, and the root node is the target variable of the query. In this graph, the edges have labels that correspond to the relation type involved in a predicate.

We illustrate this in Fig. \ref{fig:query-graph} for the example query introduced previously. In Graph Query Embedding (GQE) \cite{hamilton2018embedding}, the authors note that this graph can be employed to define a computation graph that starts with the embeddings of the entities at the leaves, and follows the structure of the query graph until the target node is reached.

GQE was one of the first models that defined a query embedding function to answer queries over KGs. The function relies on two different mechanisms, each of which handles paths and intersections, respectively. This requires generating a large dataset of queries with diverse shapes that incorporate paths and intersections.

\subsection{MPQE}

Graph Convolutional Networks (GCNs) \cite{defferrard2016convolutional,kipf2016semisupervised,gilmer2017nmp} are an extension of neural networks to graph-structured data, that allow defining flexible operators for a variety of machine learning tasks on graphs. Relational Graph Convolutional Networks (R-GCNs) \cite{schlichtkrull2017modeling} are a special case that introduces a mechanism to deal with different relation types as they occur in KGs, and have been shown to be effective for tasks like link prediction and entity classification.

In MPQE \cite{daza2020message}, the authors note that a more general query embedding function can be defined in comparison with GQE, if an R-GCN is employed to map the query graph to an embedding. The generality stems from the fact that the R-GCN uses a general message-passing mechanism to embed the query, instead of relying on specific operators for paths and intersections.

\subsection{Query2Box}

Both GQE and MPQE embed a query as a single vector (i.e., a point in space).
Query2Box~\cite{ren2020query2box} deviates from this idea and uses a box shape to represent a query.
The method further narrows the allowed embedding shape to axis-aligned hyper-rectangles. 
We will discuss more in section \ref{sec:boxes} why that is beneficial.
This method has several benefits, especially for conjunctive queries; for these queries, the answer set can be seen as the intersection of the answers to the conjuncts. 
Such an operation can be imagined with an embedded volume, but not with a vector embedding.

While this method would have made it possible to create a binary classifier, the model is not specifically trained, nor evaluated for multiple answers.

\subsection{Complex Query Decomposition}

Complex Query Decomposition (CQD) \cite{arakelyan2021complex}, is a recently proposed method for query answering based on using simple methods for 1-hop link prediction to answer more complex queries. In CQD, the link predictors used are DistMult \cite{yang2015distmult} and ComplEx \cite{trouillon2016}. Such link predictors are more data efficient than the previous methods, since they only need to be trained with the set of observed triples. In contrast, to be effective the previous methods require mining millions of queries covering a wide range of structures.

In CQD, a complex query is decomposed in terms of its binary predicates. The link predictor is  used to compute scores for each of them, and the scores are then aggregated with t-norms, which have been employed in the literature as continuous relaxations of the conjunction and disjunction operators~\cite{DBLP:journals/corr/SerafiniG16,minervini2017adversarial,vankrieken2020analyzing}.

CQD provides an answer to the query by providing a ranking of entities based on the maximization of the aggregated scores. Therefore, the evaluation procedure for CQD is the same as the previous methods.

\section{From Ranking Metrics to Actual answers}
\label{sec:rank_vs_binary}

As discussed above, there are merits to returning a hard answer set as opposed to returning a ranking. One way to obtain such binary classifications is to define a threshold within a ranking.
As we will further describe in section \ref{sec:boxes}, one can create such a threshold by using shapes (e.g. axis aligned hyper-rectangles) for query embeddings.

\subsection{Closed-world assumption}

Binary classification does introduce new challenges. 
One such challenge can be seen in the definition of a loss function that can act differently for entities within the set and entities not in the set. 
Since the knowledge graph may contain missing edges, the retrieved target set may be a subset of the ground truth. This in turn could result in entities being incorrectly used within the loss function (i.e. an incorrect closed-world assumption).

However, this is not necessarily problematic. 
We define $\mathcal{T}$ to be the ground truth target set of a query and $\mathcal{T}^\prime$ to be the retrieved target set (i.e. when directly querying the KG). 
Assuming the number of entities missing from $\mathcal{T}^\prime$ is considerably smaller than $\mathcal{V} - \mathcal{T}$, most entities that do not belong in $\mathcal{T^\prime}$ are also not answers to the query (i.e. not in $\mathcal{T}$). This means that if we sample a relatively small subset of the inverse found target set ($\mathcal{V} - \mathcal{T^\prime}$) it will likely not contain entities that are also in $\mathcal{T}$.\\
In the case where we need to be certain that our sample from $\mathcal{V} - \mathcal{T}^\prime$ does not contain entities in $T$ we could restrict our sampling process to entities which could never appear in $\mathcal{T}$. This is possible for example, by sampling entities which are incompatible with the domain and range of specific relations in a query (e.g. house entities will never appear in a \texttt{has\_sibling(a,b)} relation). Potential downsides of such methods include a potential slow down during learning or a limit in the model's overall performance, as having very different entities in $\mathcal{T}$ and our sample from $\mathcal{V} - \mathcal{T}^\prime$ could prevent our model from learning the differences between the two sets. 
On the other hand, if these two sets are very similar the model would be forced to uncover differences even when they are not very apparent. 
In fact, it is often good practise to use so-called ``hard" negative samples, which are similar to entities in $\mathcal{T^\prime}$. A better alternative for finding entities not in $\mathcal{T}$ would be using more advanced techniques as proposed in \cite{safavi2020evaluating}. 

\subsection{From ranking to classification}

Another focal point where binary classification differs from ranking as a metric, is in the way performance is measured (e.g. F-score against Mean Reciprocal Rank). On binary classification, a common performance measure would be the F-score, which is the harmonic mean between Precision and Recall, while in a ranking setting we encounter the Mean Reciprocal Rank.\\
While these metrics differ significantly, there are ways for them to relate. This insight can be evident, considering that rankings could be turned in binary classifications, using a threshold. In particular, we notice that ranking metrics typically focus on having entities in $\mathcal{T}^\prime$ higher in the rank. As a result, having many high-ranking entities that are not in $\mathcal{T}^\prime$ is also penalised. Effectively these measures then provide some notion of how well $\mathcal{T}^\prime$ and $\mathcal{V} - \mathcal{T}^\prime$ can be separated. This means that in the case of a low ranking measure, the binary classification can also under-perform. Moreover, it could either result in low precision, recall or both, depending on where the threshold is placed among the ranking.\\
Geometrically, there is also a correspondence between a ranking with a cutoff point and a system where all answer embeddings withing a given distance would be included as answers. 
One could view a classifier with high precision and low recall as having an embedding with relatively small volume, while viewing a classifier with high recall and low precision as having an embedding with relatively large volume instead. In this setting, the interpretation of a ranking measure would be whether entities in $\mathcal{T^\prime}$ are closer to our geometric query embedding than entities not in $\mathcal{T^\prime}$. This measure of closeness is defined via a distance metric (e.g. the L1 norm) and can be used in the loss function \cite{ren2020query2box}.

\section{Using Axis-aligned Boxes for Query Embedding}
\label{sec:boxes}

As discussed in section \ref{sec:methods} an entity is a valid answer to a specific structured query if it satisfies the query. 
The ultimate aim is to find the set of all valid answers, as entities in the Knowledge Graph, that satisfy the given query even when a missing edge in the KG is required for the binary predicates. 
As discussed, we could either attempt to use a cut-off point in the ranking to obtain a binary classifier, or we could train the embedding model such that it indicates a volume in the embedded space that contains the answers.
In this section we present a first possible design of such a system to show the feasibility.
We alter the earlier work done on query2box~\cite{ren2020query2box} method in two ways.
First, we do interpret the boundaries of the hyperrectangle used for the embedding as a bounding box.
All entities within the box are predicted answers to the query, while answers outside are predicted to not be answers.
Second, we do not use the embedding procedure proposed in query2box, but rather perform the embedding using the technique devised in MPQE.

\begin{figure}
	\centering
	\includegraphics[width=180pt]{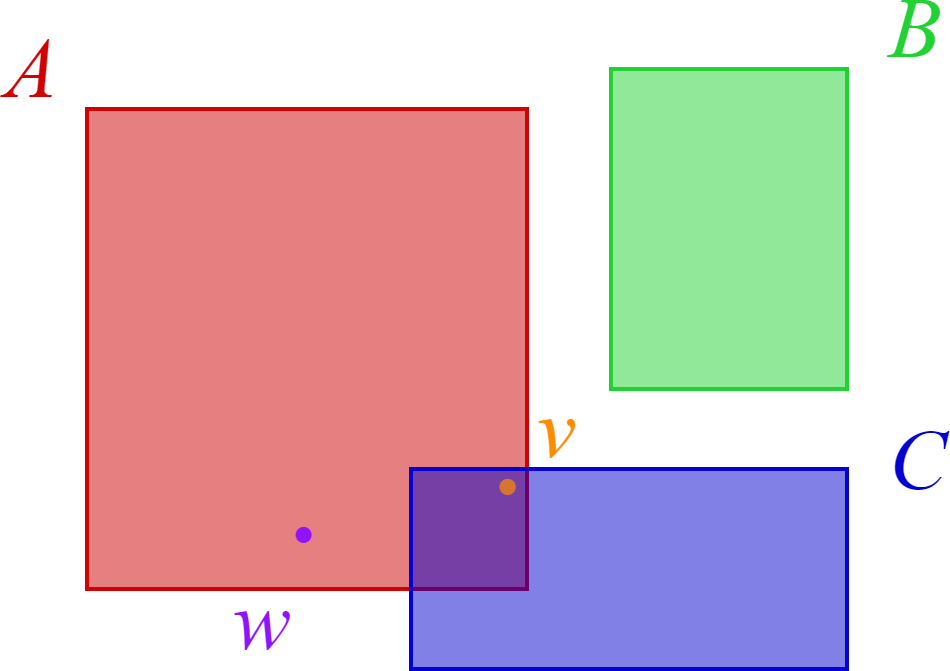}
	\caption{A small 2D query box embedding: Here there are three queries \textit{A}, \textit{B} and \textit{C}, and two entities \textit{v} and \textit{w}. In this case \textit{v} is an answer to \textit{A} and \textit{C}, whilst \textit{w} is only an answer to \textit{A}. (source \cite{thesisRuud})}
	\label{fig:query_boxes}
\end{figure}

Now, we could choose to embed entities using points, as is done in other query embedding methods. 
Then, entities that get embedded inside the box would be seen as answers to the query, while points outside of it would be seen as non-answers. This is illustrated in figure \ref{fig:query_boxes}

\begin{figure}
	\centering
	\includegraphics[width=180pt]{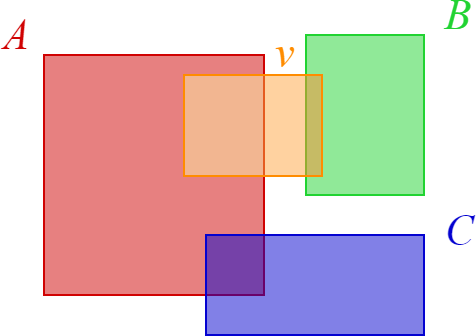}
	\caption{A small 2D query and entity box embedding: Here there are three queries \textit{A}, \textit{B} and \textit{C}, and one entity \textit{v}. In this case \textit{v} is an answer to \textit{A} and \textit{B}, but not to \textit{C}. (source \cite{thesisRuud})}
	\label{fig:query_and_entity_boxes}
\end{figure}

But, as we will discuss in more detail in the following subsection, we can also use hyper-rectangles for these. 
The choice we make in the experiments in this paper is to consider an entity, embedded as a box, to be valid answer to the query if there is an intersection between the two boxes. This is also illustrated in figure \ref{fig:query_and_entity_boxes}, for the two-dimensional case.
An alternative choice could be to consider an entity and answer in case the entity box is completely inside the query box.

To formalize this, we operate on the embedding space $\mathbb{R}^d$. 
What we want is to describe an axis-aligned hyper-rectangle in this space. 
We do this by keeping two vectors, one to indicate the center of the box and one to indicate the offset of the sides of the box.
So, in the described model every entity $v \in V$ has an embedding $\mathbf{e}_{v} \in \mathbb{R}^{2d}$.
Additionally an embedding for the query is defined that maps the full vector of the query: $\mathbf{q}\in\mathbb{R}^{2d}$. 

The boxes in $\mathbb{R}^d$ corresponding to the $2d$-dimensional vectors are defined as $p = (\mathit{Cen}(p), \mathit{Off}(p)) \in \mathbb{R}^{2d}$:
\begin{equation}
\mathit{Box}_{p}=\{ v \in \mathbb{R}^{d}:
\mathit{Cen}(p) - \mathit{Off}(p) \preceq v \preceq \mathit{Cen}(p) + \mathit{Off}(p) \},
\label{eq:box_def}    
\end{equation}
where $\preceq$ denotes element-wise inequality.

Note that a completely analog definition could be made by keeping two extreme counterpoints of the box rather than a center and offset. 

\subsection{Boxes for Entities}

\begin{figure}
    \centering
    \includegraphics[width=150pt]{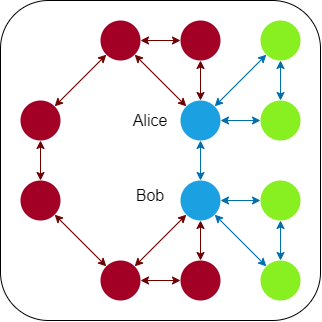}
    \caption{Here Alice and Bob are closely related in context of a specific relations (1 relation minimum), but they are not very closely related in other context (5 hops minimum). (source \cite{thesisRuud})}
    \label{fig:different_contexts}
\end{figure}

\begin{figure}
    \centering
    \includegraphics[width=150pt]{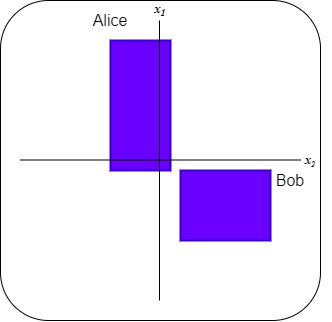}
    \caption{Here Alice and Bob are have relatively close points (seen near the origin), but also very distant points. (source \cite{thesisRuud})}
    \label{fig:different_contexts_boxes}
\end{figure}

It was already mentioned in the previous section that we represent our entity embeddings with boxes, as well.
This idea comes forward from the fact that entities could play different roles in different contexts. For example, we could have a person who both works at a university, buy is also a member of a political party. 
Having a single point to represent that person forces a query asking for members of that political party and a query asking for people working at that university to overlap. 
If we instead use a box for the entity, the query embeddings do not have that additional problem.
The issue is also illustrated in figure \ref{fig:different_contexts} and \ref{fig:different_contexts_boxes}. The nodes representing Alice and Bob are close to each other in the one context, but far away in the other one. 
In the embedding of the entities in fig. \ref{fig:different_contexts_boxes} shows that with boxes it is possible to have the entities close to each other and far away from each other at the same time.
With the entities as boxes, we can have it as an answer to two disjoint queries as illustrated in fig. \ref{fig:query_and_entity_boxes}.

\section{Proof of Concept}

In this section, we perform an evaluation of the system we discuss above. 
Note that our goal is not to provide state-of-the-art results. 
Firstly, this is because what we propose is just a proof of concept for an approximate embedding system which can find a set of answers for a query.
But, the main reason we cannot really compare with other systems is because they are evaluated with ranking metrics as discussed in section \ref{sec:rank_vs_binary}.

\subsection{Experimental setup}

\definecolor{mygray}{RGB}{153,153,153}
\definecolor{mygreen}{RGB}{00,99,00}
\definecolor{myblue}{RGB}{33,99,255}

\begin{figure}
    \centering
    \includegraphics[width=0.4\textwidth, angle=0]{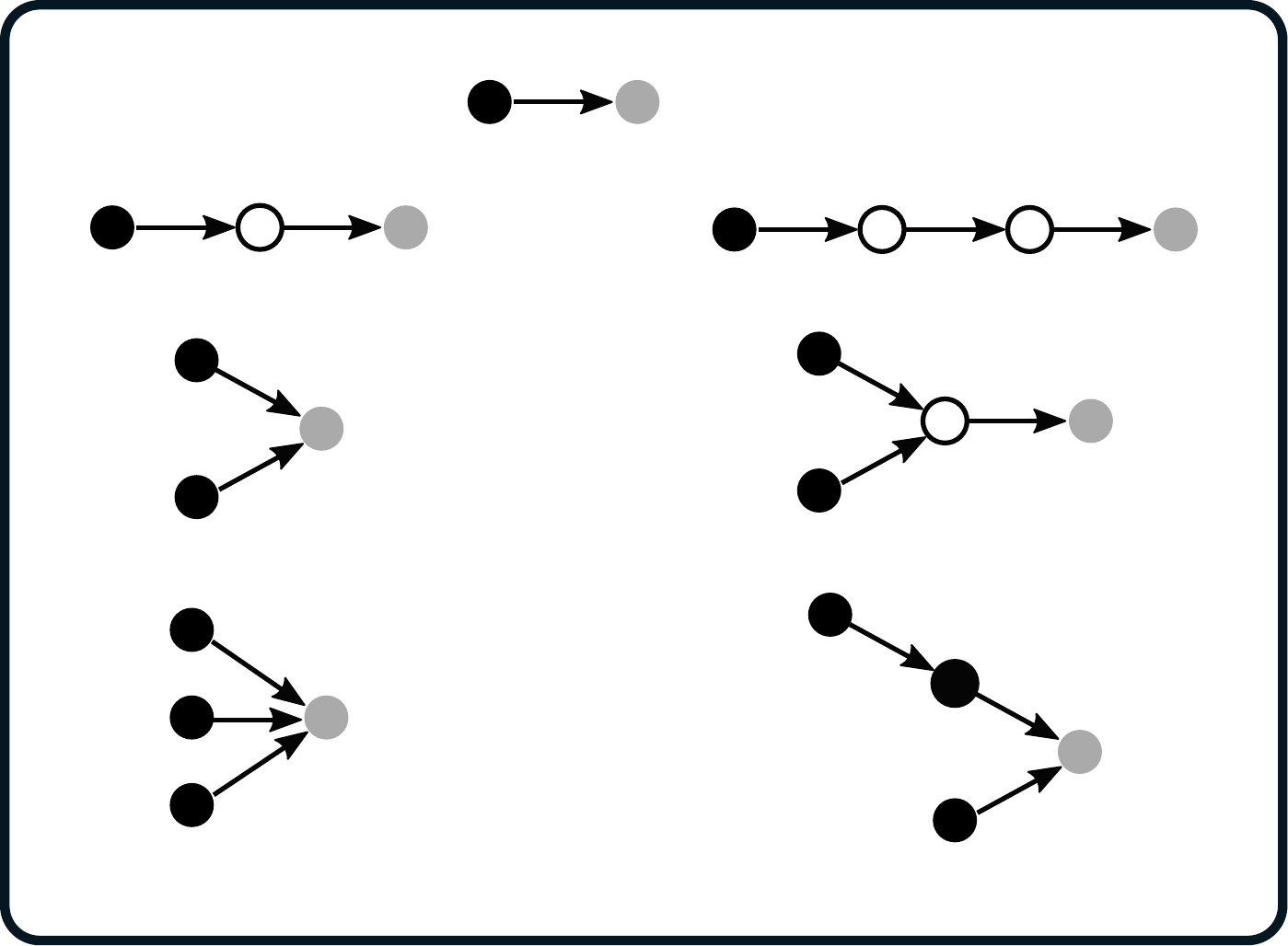}
    \caption{Used query structures for evaluation on query answering. Black nodes correspond to anchor entities, hollow nodes are the variables in the query, and the gray nodes represent the targets (answers) of the query.  (source \cite{daza2020message})}
    \label{fig:query_structures}
\end{figure}

Figure \ref{fig:query_structures} shows seven distinct query graph structures. We only consider these structures when training and testing our model for the query answering task. 
These structures were originally proposed in GQE~\cite{hamilton2018embedding}.
Each of these structures starts with actual entities from a graph (i.e. anchor entities) and ends with a set of target entities. Some of these structures are chains without any intersections (e.g. $B. \exists A, B: \text{knows}(\text{Alice},A) \land \text{is\_related\_to}(A,B)$), whilst other only have intersections (e.g. $B. \exists B: \text{knows}(\text{Alice},B) \land \text{is\_related\_to}(\text{Bob},B)$) or even combinations of both. 
Our goal is to train a model that finds the answer set of a given query, using a query embedding. 
This is in contrast to other related work \cite{ren2020query2box,hamilton2018embedding,daza2020message} as we want to be able to find multiple answers. As mentioned before, we could create such a set by embedding the query as box, thus getting a hard boundary for separating entities in and not in the target set.

\subsubsection{Datasets}
While previous work \cite{daza2020message,hamilton2018embedding} incorporated multiple datasets, our implementation has yet solely been tested on the AIFB dataset. This dataset is a knowledge graph of academic institution in which persons, organizations, projects, publications, and topics are the entities. Table \ref{tab:stats} give some statistics of this dataset and also for two more datasets often used for the evaluation of approximate query answering.

\begin{table}
   \caption{Statistics of the knowledge graphs that were used for training and evaluation.}
    \label{tab:stats}
    \vskip 0.15in
    \centering
    \resizebox{0.49\textwidth}{!}{
        \begin{tabular}{lrrrr}
           \toprule
			& AIFB    & MUTAG  & AM        \\
			\midrule
			Entities       &  2,601  & 22,372 &   372,584 \\
			Entity types   &      6  &      4 &         5 \\
			Relations      & 39,436  & 81,332 & 1,193,402 \\
			Relation types &     49  &      8 &        19 \\
			\bottomrule
			\hfill

        \end{tabular}}
\end{table}

\subsubsection{Query Generation}
To train our model we have to sample for query graphs from our dataset. This is done by initially sampling anchor nodes and relations which are later used to form graphs based on specific query patterns (fig. \ref{fig:query_structures}).

After acquiring the anchor nodes and the relations connecting them, we can obtain the target set. Although this may appear straightforward, there are some caveats. The biggest one is that some queries contain considerable sets of potential target entities (over 100,000 answers). Because we sample for edges first these particular graphs actually appear often.

Luckily, for most query structures this was not the case, but specifically the 2-chain and 3-chain query structures occasionally suffer from it. This is likely explained by the fact that knowledge graphs contain ``hub nodes", nodes with a very high degree, to which a plethora of other nodes connect via a certain relation. 
Table \ref{tab:avg_targets} shows the average size of the target sets of sampled queries for the aforementioned datasets. One interesting thing to note is that for the AM dataset the 3-chain-inter structure actually had the largest average target set. This could indicate that this problem is indeed very graph-dependent. Since this is a problem with the AIFB dataset, we limit the query target sets to a maximum of 100 answers.\\
We also sample for entities not in the target set to be used as negative samples during training. For the query structures that contain an intersection we incorporate hard negative samples by finding entities that would have been in the target set if the conjunctive intersections were to be relaxed to disjunctions.

\begin{table}
    \vskip 0.15in
    \centering
    \caption{Average number of multiple answers to different queries structures, across the used datasets. (results were earlier reported in \cite{thesisTeodor})}
\label{tab:avg_targets}
    \resizebox{0.48\textwidth}{!}{
        \begin{tabular}{lcccccccc}
            \toprule
                         &   \multicolumn{2}{c}{AIFB} &   \multicolumn{2}{c}{MUTAG} &   \multicolumn{2}{c}{AM} \\
            Structure    &   Train     &     Test      &   Train      &     Test    &   Train     &       Test      \\  
            \midrule                                                                                                              
            1-chain   	 &     3.4     &      1.2      &     1.9      &     1.1      &     1.2     &      1.0       \\ 
            2-chain		 &    34.5     &      6.4      &    13.4      &     4.7      &    10.2    &      3.5     \\ 
            3-chain      &\bf 47.0     &\bf   7.2      &\bf 17.6      &\bf  5.4      &    13.8    &      3.7      \\ 
            \midrule                                                                                                              
            2-inter      &     9.3     &      3.2      &     1.6      &     1.3      &    9.1     &      3.5     \\ 
            3-inter      &     5.1     &      2.8      &     1.0      &     1.0      &    7.4     &      2.9      \\ 
            \midrule
            3-inter-chain&    15.5     &      4.2      &     1.9      &     1.7       &    10.3    &      3.5      \\
            3-chain-inter&    22.8     &      5.6      &     2.6      &     2.3       &\bf 15.2    &\bf   4.4        \\ 
            \bottomrule
            \hfill
        \end{tabular}}
\end{table}

\subsubsection{Evaluation}

In order to test whether the model is actually able to find answers to queries that involve edges which are not in the graph, careful preparation of our data splits was necessary. 
We started by our original graph and marked 10\% of the edges to be removed (they are still there at this stage).
Then, we sample the graph for the query patterns.
If the sample makes use of any edge marked as removed, it will be added to either the validation set or the test set (10/90 split).
If the sample contains no such marked edge, then we put it in the training set.
This way, we end up with validation and test queries that make use of at least one edge that is not in the graph seen during training.

Post sampling, we end up with around 2 million targets and the corresponding query graphs to be used in the training set. 
For the validation set we used about 30,000 targets worth of queries and for the test set we will had approximately 300,000 targets worth of query graphs. The validation set is also used to perform early stopping in case specific conditions were not met.

\begin{figure}
    \centering
    \includegraphics[width=0.5\textwidth, angle=0]{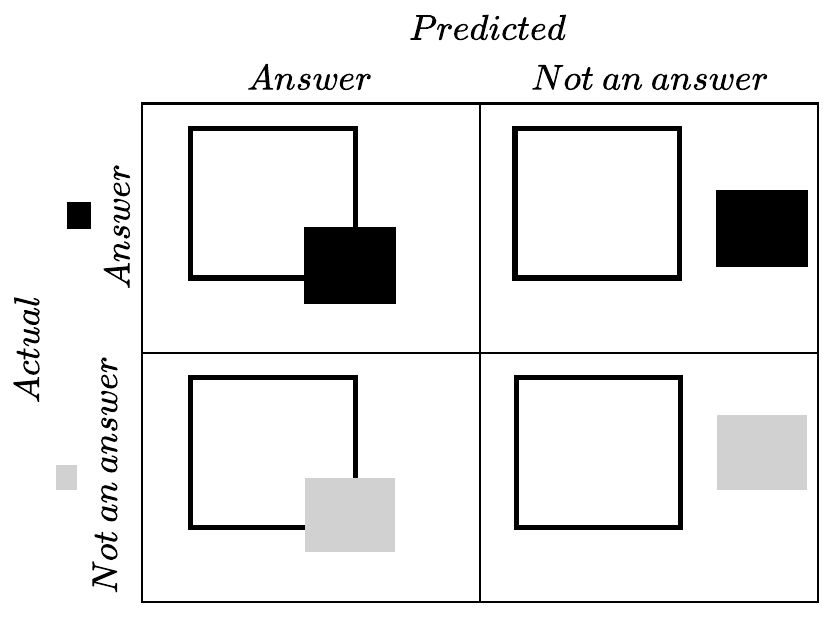}
    \caption{Model of the confusion matrix used for evaluation of the results, the empty box is representation of a query, the black and the gray box are respectively a valid and a invalid answer to the query. (source \cite{thesisTeodor})}
    \label{fig:confusion_matrix}
\end{figure}

Since our method uses boxes, which allow for binary classification, we report our model's performance in the form of a confusion matrix (see figure \ref{fig:confusion_matrix}). Given the fact that our entities are also boxes, we have more freedom to choose when an entity is considered an answer. 

This is because entities now inhabit more space than a single point which allows for partial overlap with query boxes. In order to allow flexibility we have decided that an entity is considered an answer to a query if its box representation overlaps with the box representation of the respective query box. Naturally, other more strict conditions could be applied such as requiring full overlap or define a fraction based threshold (e.g. requiring at least 50\% overlap). We expect these conditions to change based on the potential downstream task.

\definecolor{myyellow}{RGB}{227, 200, 0}
\definecolor{mygreen}{RGB}{151, 208, 119}
\definecolor{myred}{RGB}{241, 156, 153}

\subsubsection{Model}

\begin{figure}
	\centering
	\noindent\makebox[\textwidth]{\includegraphics[width=0.98\textwidth, angle=0]{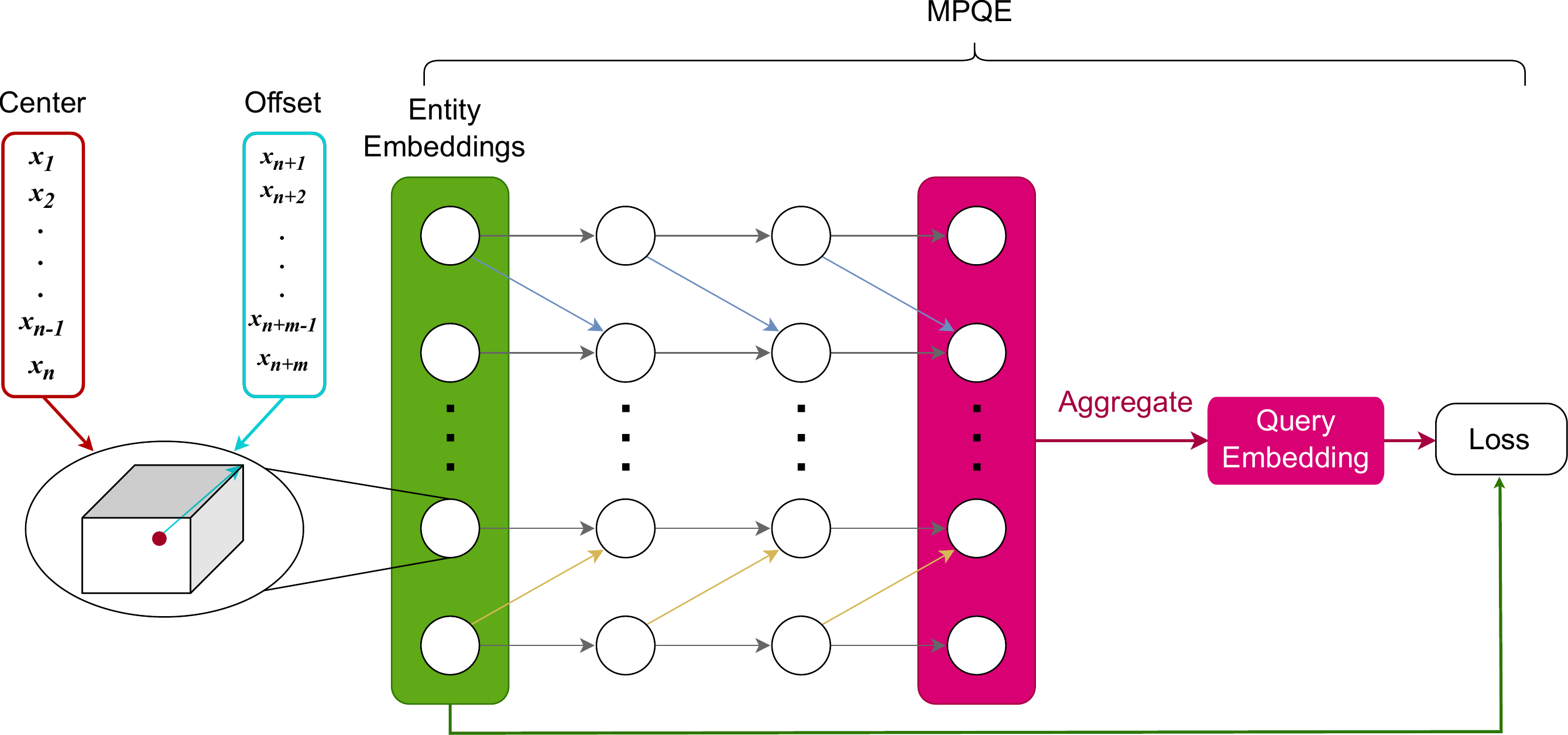}}
	\caption{The MPQB model used in this proof of concept.  (adapted from a figure in~\cite{thesisRuud}) }
	\label{fig:box_model}
\end{figure}

Our model has the same basic functionality as the MPQE \cite{daza2020message} model. 
MPQE is used as an embedding component, but the input and output are interpreted as boxes. 
MPQE first performs several steps of message passing using an R-GCN architecture after which the node states are aggregated to form the query embedding. With this query embedding a loss function is evaluated which is used as a signal (using SGD) to update the embeddings and weights in the network.
For the aggregation operation we have several options (\textit{SUM}, \textit{MAX}, \textit{TM}, \textit{MLP}) at the end of our model. We test our model with some of these different aggregation functions.

Since we train an embedding matrix (as opposed to having a latent embedding to start with) we need to initialize it. 
We do this by sampling the 32 dimensional center vectors from a uniform distribution between 0 and 10, whilst sampling the 32 dimensional offset vectors from a unit Gaussian with a mean 3.

For TM aggregation, the MPQE model uses 3 layers; the TM aggregation function requires a number of message passing steps equal to the query diameter, in our case 3. 
For the MLP aggregation function we applied a two layer fully-connected MLP. 
As for the non-linearities in our model, we used the ReLU function.
To update the parameters of the model we used Adam optimizer with a learning rate of 0.01.

Our code base is based on PyTorch. In particular, we made use of the library PyTorch Geometric \cite{fey2019fast}, which is a PyTorch extension specialised for graph-based models. While there are potential baselines to consider \cite{hamilton2018embedding,daza2020message}, they are not suitable for our work. 
This happens because we perform a binary classification as opposed to ranking-based methods. 
To our knowledge there have not been any related work that performed binary classification in the context of approximate graph querying. 
In the area of link prediction, we do find some work, like the early work on Neural Tensor Networks \cite{NIPS2013_b337e84d} and a more recent one which looks at triple classification \cite{dong2019triple}.
This did not prove to be a major concern, as our main goal was not to achieve state-of-the-art results, but rather explore whether this direction of research may prove worthwhile.

\subsection{Results}

After having trained the MPQB model for over 200,000 iterations it appeared to still not have converged. 
After this amount of iterations the query boxes seemed to not overlap with any target boxes (i.e. no entities in $\mathcal{T}^\prime$ were returned). 
Apart from training the model for longer and on multiple epochs, there are some other settings that could still be experimented with. 
For example, how many samples are in each epoch (less samples allow for training on more epochs), whether we use $\mathcal{T}^\prime$ fully during train or use a subset, and how many entities should be in our sample from $\mathcal{V} - \mathcal{T}^\prime$. The latter two settings also influence how many distinct queries we could train on within a given time span. In may be worth noting that previous works \cite{hamilton2018embedding,daza2020message,ren2020query2box} train using single positive samples. While we want to focus on answering queries with multiple answers, we do not necessarily need to train on multiple answers. In theory, if a method can produce a good ranking, it should also be able to produce a good classification, given that the optimal thresholds for these rankings could be found.\\
Since we do not have direct result in a manner we would have liked, we will instead analyse the trained models to see if there are relevant insights to be found. For this we looked at models using different aggregation functions, trained on the AIFB dataset.\\
While we have no intersections between query boxes and target boxes, we could still look whether the target boxes (from $\mathcal{T}^\prime$) appear relatively close to the entity boxes, when compared to the box representations of entities in $\mathcal{V} - \mathcal{T}^\prime$. This effectively provides some measure as to whether the produced rankings are good. Table \ref{tab:results} shows these results. While these scores may not indicate state-of-the-art results, they do seem to suggest that the model did at least produce decent non-trivial rankings using the SUM and TM aggregators. This could suggest that further research is indeed in order. The fact that TM outperformed SUM is not surprising considering that it is a more involved method that also takes query diameter into account. This result is also in line with the findings in \cite{daza2020message}. A more surprising result is that the MLP method did not seem to perform well at all. This could be a result of a faulty implementation, or an implementation that simply does not work for boxes as is. Overall, the results seem promising.

\begin{table}
\caption{Percentage (\%) of answers embedded closer to the query box compared to a non answer, with regard to the query structure, using different aggregation function. Tested on AIFB dataset. (results were earlier reported in \cite{thesisTeodor})}
\label{tab:results}
\vskip 0.15in
\centering
\resizebox{0.45\textwidth}{!}{
\begin{tabular}{lcccccccc}
\toprule
             &   \multicolumn{2}{c}{AIFB}  \\
Structure    &     SUM       &     TM      &   MLP      \\  
\midrule                                                                                                              
1-chain   	 &     67.48     &\bf  69.84   &     0.0       \\ 
2-chain		 &     68.78     &\bf  75.85   &     0.0         \\ 
3-chain      &     76.55     &\bf  79.86   &     0.0          \\ 
\midrule                                                                                                              
2-inter      &     62.09     &\bf  63.10   &     0.0         \\ 
3-inter      &     63.32     &\bf  63.35   &     0.0         \\ 
\midrule
3-inter-chain&     67.61     &\bf  67.91   &     0.0          \\
3-chain-inter&     68.87     &\bf  72.43   &     0.0        \\ 
\bottomrule
\end{tabular}
}
\end{table}

\section{Conclusion and Outlook}

In this work, we looked critically at the currently prevailing evaluation strategy for approximate complex structured query algorithms for knowledge graphs.
Typically, these systems take a query as an input and produce a ranking of all entities in the KG as an output.
The performance of these systems is than determined using metrics typically used in information retrieval.

What we propose is to augment the current evaluations by also requiring these systems to produce a binary classification of the nodes into a class of answers and one of non-answers.
This is needed because many applications can simply not work with a ranking and need a fixed set of answers to work with.

As a first proof of concept, we have adapted ideas from MPQE and query2Box, and created an embedding algorithm that represents the queries and the entities as axis-aligned hyper-rectangles.
We noticed that the performance of this system is pretty low, and expect that future works can heavily improve upon this first attempt.

As future research directions, we see a need to expand our experiments to include other query types (disjunctions, negations, filters, etc. ), in order to show the generalizability of our approach. 
This will, however, require new representation for the volumes as these operations are not possible if we would stay with just boxes. For example, the negation of a box, would no longer be a box.

Moreover, we it needs to be investigated how our method can be applied on different kinds of graphs. This will give us insights as to what changes need to be made in terms of training data (via query generation) as well as the effects on model performance. 
Also, it seems worth experimenting with different geometric representations for the parts of the query (anchor, variables and targets). Finally, since our experiments were relatively small-scale, further research could also start by simply experimenting with different settings for our current architecture.

\bibliographystyle{splncs04}
\bibliography{refs}

\begin{thebibliography}{10}
\providecommand{\url}[1]{\texttt{#1}}
\providecommand{\urlprefix}{URL }
\providecommand{\doi}[1]{https://doi.org/#1}

\bibitem{thesisTeodor}
Aleksiev, T.: Answering approximated graph queries, embedding the queries and
  entities as boxes (2020), {BSc.} thesis, Computer Science, Vrije Universiteit
  Amsterdam, Supervised by Cochez, M.

\bibitem{arakelyan2021complex}
Arakelyan, E., Daza, D., Minervini, P., Cochez, M.: Complex query answering
  with neural link predictors. In: International Conference on Learning
  Representations (2021), \url{https://openreview.net/forum?id=Mos9F9kDwkz}

\bibitem{thesisRuud}
van Bakel, R.: Box {R-GCN}: Structured query answering using box embeddings for
  entities and queries (2020), {BSc.} thesis, Computer Science, Vrije
  Universiteit Amsterdam, Supervised by Cochez, M.

\bibitem{daza2020message}
Daza, D., Cochez, M.: Message passing query embedding. In: {ICML Workshop -
  Graph Representation Learning and Beyond} (2020),
  \url{https://arxiv.org/abs/2002.02406}

\bibitem{defferrard2016convolutional}
Defferrard, M., Bresson, X., Vandergheynst, P.: Convolutional neural networks
  on graphs with fast localized spectral filtering. In: Advances in neural
  information processing systems. pp. 3844--3852 (2016)

\bibitem{dong2019triple}
Dong, T., Wang, Z., Li, J., Bauckhage, C., Cremers, A.B.: Triple classification
  using regions and fine-grained entity typing. In: Proceedings of the AAAI
  Conference on Artificial Intelligence. vol.~33, pp. 77--85 (2019)

\bibitem{fey2019fast}
Fey, M., Lenssen, J.E.: Fast graph representation learning with {PyTorch}
  {Geometric}. arXiv preprint arXiv:1903.02428  (2019)

\bibitem{gilmer2017nmp}
Gilmer, J., Schoenholz, S.S., Riley, P.F., Vinyals, O., Dahl, G.E.: Neural
  message passing for quantum chemistry. In: Proceedings of the 34th
  International Conference on Machine Learning, {ICML} 2017, Sydney, NSW,
  Australia, 6-11 August 2017. pp. 1263--1272 (2017)

\bibitem{hamilton2018embedding}
Hamilton, W., Bajaj, P., Zitnik, M., Jurafsky, D., Leskovec, J.: Embedding
  logical queries on knowledge graphs. In: Bengio, S., Wallach, H., Larochelle,
  H., Grauman, K., Cesa-Bianchi, N., Garnett, R. (eds.) Advances in Neural
  Information Processing Systems. vol.~31, pp. 2026--2037. Curran Associates,
  Inc. (2018),
  \url{https://proceedings.neurips.cc/paper/2018/file/ef50c335cca9f340bde656363ebd02fd-Paper.pdf}

\bibitem{hogan2020knowledge}
Hogan, A., Blomqvist, E., Cochez, M., d'Amato, C., de~Melo, G., Gutierrez, C.,
  Gayo, J.E.L., Kirrane, S., Neumaier, S., Polleres, A., et~al.: {Knowledge}
  {Graphs}. arXiv preprint arXiv:2003.02320  (2020)

\bibitem{kipf2016semisupervised}
Kipf, T.N., Welling, M.: Semi-supervised classification with graph
  convolutional networks. arXiv preprint arXiv:1609.02907  (2016)

\bibitem{vankrieken2020analyzing}
van Krieken, E., Acar, E., van Harmelen, F.: {Analyzing Differentiable Fuzzy
  Implications}. In: {Proceedings of the 17th International Conference on
  Principles of Knowledge Representation and Reasoning}. pp. 893--903 (9 2020).
  \doi{10.24963/kr.2020/92}, \url{https://doi.org/10.24963/kr.2020/92}

\bibitem{minervini2017adversarial}
Minervini, P., Demeester, T., Rockt{\"{a}}schel, T., Riedel, S.: Adversarial
  sets for regularising neural link predictors. In: {UAI}. {AUAI} Press (2017)

\bibitem{noy2019industry}
Noy, N., Gao, Y., Jain, A., Narayanan, A., Patterson, A., Taylor, J.:
  Industry-scale knowledge graphs: lessons and challenges. Communications of
  the ACM  \textbf{62}(8),  36--43 (2019)

\bibitem{ren2020query2box}
Ren, H., Hu, W., Leskovec, J.: Query2box: Reasoning over knowledge graphs in
  vector space using box embeddings (2020)

\bibitem{safavi2020evaluating}
Safavi, T., Koutra, D., Meij, E.: Evaluating the calibration of knowledge graph
  embeddings for trustworthy link prediction (2020)

\bibitem{schlichtkrull2017modeling}
Schlichtkrull, M., Kipf, T.N., Bloem, P., Van Den~Berg, R., Titov, I., Welling,
  M.: Modeling relational data with graph convolutional networks. In: European
  Semantic Web Conference. pp. 593--607. Springer (2018)

\bibitem{DBLP:journals/corr/SerafiniG16}
Serafini, L., d'Avila Garcez, A.S.: Logic tensor networks: Deep learning and
  logical reasoning from data and knowledge. CoRR  \textbf{abs/1606.04422}
  (2016), \url{http://arxiv.org/abs/1606.04422}

\bibitem{NIPS2013_b337e84d}
Socher, R., Chen, D., Manning, C.D., Ng, A.: Reasoning with neural tensor
  networks for knowledge base completion. In: Burges, C.J.C., Bottou, L.,
  Welling, M., Ghahramani, Z., Weinberger, K.Q. (eds.) Advances in Neural
  Information Processing Systems. vol.~26, pp. 926--934. Curran Associates,
  Inc. (2013),
  \url{https://proceedings.neurips.cc/paper/2013/file/b337e84de8752b27eda3a12363109e80-Paper.pdf}

\bibitem{trouillon2016}
Trouillon, T., Welbl, J., Riedel, S., Gaussier, {\'{E}}., Bouchard, G.: Complex
  embeddings for simple link prediction. In: {ICML}. {JMLR} Workshop and
  Conference Proceedings, vol.~48, pp. 2071--2080. JMLR.org (2016)

\bibitem{yang2015distmult}
Yang, B., Yih, W., He, X., Gao, J., Deng, L.: Embedding entities and relations
  for learning and inference in knowledge bases. In: 3rd International
  Conference on Learning Representations, {ICLR} 2015, San Diego, CA, USA, May
  7-9, 2015, Conference Track Proceedings (2015)

\end{thebibliography}

\end{document}